\documentclass[manuscript]{emulateapj}
\newcommand{\DxDy}

\shorttitle{Obliquity Evolution}
\shortauthors{Rogers \& Lin}

\begin{document}

\title{On the Tidal Dissipation of Obliquity}

\author{T.M. Rogers}
\affil{Department of Planetary Sciences, University of Arizona,
    Tucson, AZ, 85719}
\email{tami@lpl.arizona.edu}
\author{D.N.C. Lin}
\affil{Astronomy and Astrophysics Department, University of California, Santa Cruz, CA 95064}
\affil{Kavli Institute for Astronomy and Astrophysics and School of Physics, Peking University, China}
\email{lin@ucolick.org}

\begin{abstract}
We investigate tidal dissipation of obliquity in hot Jupiters.  Assuming an
initial random orientation of obliquity and parameters relevant to the observed population, the obliquity of
hot Jupiters does not evolve to purely aligned systems.  In fact, the
obliquity evolves to either prograde, retrograde or 90$^{o}$ orbits
where the torque due to tidal perturbations vanishes.  
This distribution is incompatible with observations which show that hot jupiters
around cool stars are generally aligned.  This calls into question the
viability of tidal dissipation as the mechanism for obliquity
alignment of hot Jupiters around cool stars.  
\end{abstract}


\section{Introduction}

Hot Jupiters (with masses comparable to that of Jupiter and orbital
periods less than a week or so) are found around 1-2 \% of solar 
type stars.  A widely adopted scenario for their origin is that 
these planets formed at several AU from their host stars and
underwent inward migration due to tidal interaction
with their natal disks \citep{lin96}.  Another class of dynamical models assumes the hot Jupiters were
relocated to close proximity to their host stars through close
encounters between planets, secular chaos, or Kozai resonance with
companion stars \citep{rf96,wm03,fab07,wu11,naoz11}.  
%Their
%eccentricities were subsequently damped by the dissipation 
%of tidal perturbation exerted on them by their host stars \citep{pt77,ivanov04}.
%Such eccentricity damping due to the dissipation of tides within the planet
%has no effect on the obliquity.  However, dissipation of tides induced
%by the planet on the star can transfer angular momentum between the
%stellar spin and planetary orbit, thus causing obliquity evolution.  

New observations are continually testing and constraining these
theories.  To date the obliquity between planets and
their host stars have been measured using the Rossiter-McLaughlin
effect in more than 50 systems.
Several hot Jupiters have been observed to have high obliquity, $\Theta$ (including
retrograde orbits in which $\Theta > \pi/2$), while many show
alignment.  In general, hot Jupiters around cool stars with
effective temperature $T_\ast <6,250$K, tend to be aligned, while those
around hot stars, $T_\ast >6,250$K, appear to be misaligned \citep{winn10}.
In order to account for this dichotomy,  \cite{winn10} and
subsequently, \cite{alb12} have suggested that 1) all hot Jupiters
were relocated to close proximity to their host stars by one of the
dynamical processes described above, resulting in a random distribution
of obliquity, 2) the obliquities of those planets around cool stars
are damped by efficient tidal dissipation in the convective envelopes of these cool stars and 3) the
obliquities of those planets around hot stars, with radiative
envelopes, reflects that of the initial random obliquity distribution
because tidal dissipation in these systems is inefficient.  

Quantitatively, \cite{alb12} evaluated the magnitude of the 
tidal dissipation time scale in stars with convective envelopes,
$\tau_{CE}$, and those with radiative envelopes, $\tau_{RA}$ using
models of equilibrium tides for convective solar-type stars  
and dynamical tides for radiative, intermediate-mass stars \citep{zahn77}. 
They showed a correlation between the magnitude of misalignment,
$\Theta$, and the dissipation timescales.
%, $\tau_{CE}/\tau_\ast$ and
%$\tau_{RA}/\tau_\ast$, where $\tau_\ast$ is the typical stellar 
%age ($\sim$ a few Gyr).

Despite this suggestive evidence, there are large uncertainties in the
equilibrium-tide model that \cite{alb12} have adopted, mostly due to a
controversial prescription for turbulent dissipation  of the tides
\citep{gn89a,terquem98,ogilvie07,barker10}.  Using
Zahn's model, \cite{alb12} calculate that the magnitudes of 
$\tau_{CE}/\tau_\ast$ and 
$\tau_{RA}/\tau_\ast$ vary by several orders of magnitude
(see their Figures 24 \& 25).  The relatively small value of 
$\tau_{CE}/\tau_\ast$ obtained for many of the aligned 
hot Jupiters around solar type stars pose a challenge to 
their retention against substantial orbital decay, 
unless the time scale for obliquity damping is substantially smaller than that for orbital decay.

In a thorough theoretical analysis, \cite{lai12} showed that
the timescale for obliquity damping could be substantially
different from that of orbital decay. He identified a component of
the tidal torque which affects the obliquity alignment but not the 
orbital decay. He showed that, for the forcing frequency of this 
particular component of the tidal perturbation, inertial waves may 
be excited to provide a dynamical tidal response which could lead 
to much more efficient energy dissipation and hence, shorter
dissipation timescales. Therefore, the \cite{lai12} 
theory appears to provide a potential mechanism to account for hot 
Jupiters whose obliquity has been damped in absence of substantial 
orbital decay.

In \S2, we briefly recapitulate Lai's theory and in \S3 show that the orbital decay paradox can only 
be resolved if the timescale for obliquity evolution is much
smaller than that for semi-major axis decay.
In the limit of negligible orbital decay, we show in \S3, that 
tidal dissipation does not lead to obliquity alignment.
Both of these effects have been clearly stated by \cite{lai12}.
Our contribution is to provide the results of numerical 
integration for various limiting cases and compare them 
with the observational data.  We discuss 
the implications of our results in \S4. 

\section{Semi major axis decay and obliquity alignment}

We first briefly recapitulate the theory of equilibrium and 
dynamical tides raised by hot Jupiters on their host stars.

\subsection{Components of planets' tidal perturbation on their 
spinning host stars} 

The tidal potential $U({\bf r}, t)$ imposed by a 
planet with mass $M_p$, with a circular orbit 
and an orbital angular frequency ${\Omega_p}$ 
on a star with a mass $M_\ast$ and a uniform 
spin with angular frequency ${\Omega_\ast}$ can 
be approximated to the lowest order in terms 
of spherical harmonics in a frame centered on 
the star with the z-axis parallel to the stellar 
spin such that 
\begin{equation}
U({\bf r}, t)= - \sum_{m m^\prime} U_{m m^\prime} (M_p, \Theta) 
r^2 Y_{2 m} (\theta, \phi) {\rm exp} \left(- i m^\prime \Omega_p t \right)
\end{equation}
where the obliquity $\Theta$ is the angle between 
the stellar spin {\bf S} and planet's orbital angular 
momentum vector {\bf L}, whereas the magnitudes are
given by $S = I \Omega_\ast$ and $L = M_p a^2 \Omega_p$, 
$I= k M_\ast R_\ast^2$, where $R_\ast$ is the stellar
radius, $k (\simeq 0.1)$ and $I$ is the moment of inertia.  In a frame 
co-rotating with the stellar spin, the forcing 
frequency is $\omega_{m m^\prime} = m^\prime \Omega_p 
- m \Omega_\ast$ with seven components contributing 
to obliquity evolution \citep{bog09}.  

%Using a common 
%constant lag time $t_{m m^\prime}$ model for all 
%seven components, \cite{lai12} derived the standard 
%torque formula for the equilibrium tide derived by 
%\cite{alex73,hut80} with an equivalent quality
%factor $Q_a$ (see Eqs. 2-4).  

\cite{lai12} pointed out that it is possible 
for some components of the planets' tidal perturbing 
potential to have sufficiently small $\omega_{m m^\prime} 
< 2 \Omega_{\ast}$ to allow the corilois effect to 
provide the necessary restoring force for the excitation 
of inertial waves \citep{greenspan}.  In cool stars these waves lead to dynamical tides \citep{ogilvie07}.  
For some forcing frequencies, the inertial waves may 
converge onto attractors and dissipate more efficiently 
\citep{ogilvie04}.  \cite{lai12} showed that in the 
limit of small $\Omega_\ast (< \Omega_p)$, the only 
component of the perturbing potential with sufficiently 
small $\omega_{m m^\prime}$ is that associated with 
$(m, m^\prime) =(1, 0)$.  For this component, tidal 
dissipation can lead to a shorter dissipation timescale and obliquity
evolution without significant orbital decay.  For more rapidly 
spinning stars (with $\Omega_\ast > \Omega_p/2$), 
it is possible to excite other components of the 
tidal response which would lead to both obliquity 
evolution and orbital decay. 

\subsection{Evolutionary equations}
We first consider the possibility that the dissipation timescale, $t_{m m^\prime}$
is indentical for all $(m,m^\prime)$ components.  In this  
case, dissipation of the equilibrium tide leads to 
\begin{equation}
{{\dot a}_e \over a} = - { 1 \over \tau_e} \left( 1 - {\Omega_\ast
\over \Omega_p} {\rm cos} \Theta \right)
\label{eq:ae}
\end{equation}
\begin{equation}
{{\dot \Omega}_{\ast e} \over \Omega_\ast} = { 1 \over \tau_e} 
\left({ L \over 2 S} \right) \left[ {\rm cos} \Theta - \left(
{\Omega_\ast \over 2 \Omega_p} \right)  ( 1 + {\rm cos}^2 \Theta) \right]
\label{eq:omegae}
\end{equation}
\begin{equation}
{\dot \Theta}_e =- { 1 \over \tau_e} 
\left({ L \over 2 S} \right) {\rm sin} \Theta \left[  1 - \left(
{\Omega_\ast \over 2 \Omega_p} \right)  \left( {\rm cos} \Theta - 
{S \over L} \right) \right]
\label{eq:thetae}
\end{equation}
where $\tau_e = (Q_a / 3 k_2) (M_\ast/M_p) (a /R_\ast)^5 (P/2 \pi)$
is the characteristic orbital evolution time scale, $k_2$ is the 
Love number, and $Q_a$ is the highly uncertain quality factor.
In addition to turbulent dissipation of equilibrium tides, $Q_a$ 
may also include contributions from the dissipation of internal gravity 
and inertial waves in solar type stars \citep{ogilvie07}.  
For rapidly spinning stars, the dynamical response associated with
the inertial waves may lead to the evolution of $a$ and $\Theta$
on similar time scales.

However, for slowly spinning stars (with $\Omega_\ast < \Omega_p/2$), 
only the forcing frequency associated with the $(m, m^\prime)=(1, 0)$
component can excite the inertial modes, leading to a dynamical 
response, and relatively short timescales $\tau_{10} ( < <
\tau_e)$.  The dynamical tide's contribution associated with the (1,0) component 
of the torque leads to rates of change of $a$, $\Omega_\ast$, and $\Theta$
such that
\begin{equation}
{{\dot a} \over a} = {{\dot a}_e \over a} ,
\label{eq:at}
\end{equation}
\begin{equation}
{{\dot \Omega}_{\ast} \over \Omega_\ast} = 
\left({{\dot \Omega}_{\ast} \over \Omega_\ast}\right)_e +
\left({{\dot \Omega}_{\ast} \over \Omega_\ast}\right)_{10} +
\left({{\dot \Omega}_{\ast} \over \Omega_\ast}\right)_{10, e},
\label{eq:omegat}
\end{equation}
\begin{equation}
{\dot \Theta} = \left({\dot \Theta} \right)_e +
\left({\dot \Theta} \right)_{10} - \left({\dot \Theta} \right)_{10, e} ,
\label{eq:thetat}
\end{equation}
where
\begin{equation}
\left({{\dot \Omega}_{\ast} \over \Omega_\ast}\right)_{10} = -
{ 1 \over \tau_{10}} \left( {\rm sin} \Theta {\rm cos} \Theta  \right)^2
\label{eq:omega10}
\end{equation}
\begin{equation}
\left({\dot \Theta} \right)_{10} =-
{ 1 \over \tau_{10}} {\rm sin} \Theta {\rm cos}^2 \Theta  
\left( {\rm cos \Theta + {S \over L} }\right),
\label{eq:theta10}
\end{equation}

\begin{equation}
{{\dot \Omega}_{\ast, 10, eq} \over {\dot \Omega}_{\ast, 10}} =
{{\dot \Theta_{10, eq} \over {\dot \Theta}_{10}} =
{Q_{10} \over Q_a } {k_2 \over k_{10}}},
\label{eq:omega10e}
\end{equation}

\begin{equation}
\tau_{10} = \left( {4 Q_{10} \over 3 k_{10}} \right) 
\left( { M_\ast\over M_p} \right) \left( {a \over R_\ast}
\right)^5 \left( {S \over L} \right) \left({ P\over 2 \pi} \right)
\label{eq:tau10}
\end{equation}
such that $\tau_{10}/\tau_e \sim (Q_{10} / Q_a)(k_2 / k_{10}) (S/L)$.
Lai (2012) pointed out that since the $(m,m^\prime)=(1,0)$ component does not 
contribute to $\dot a$, obliquity alignment can occur, in principle, 
prior to any significant orbital decay if $\tau_{10} < < \tau_e$.  

\section{Computational results}
\subsection{Evolution of Semi-Major axis and Obliquity due to
  Equilibrium Tide}

Here we show that the obliquity alignment due to the equilibrium tide
is accompanined by substantial orbital evolution.  We first neglect any extra 
contribution from the $(m, m^\prime) = (1,0)$ component by numerically 
integrating Equations (\ref{eq:ae}-\ref{eq:thetae}), with 
a fourth order Runge-Kutta scheme, for an inital value of
$\Theta_{o}=45^{o}$.  If we fix $k
(R_{\ast}/a_{o})^{2}(M_{\ast}/M_{p})=1$, we have a
set of solutions which depend only on the initial value of S/L.  The
solutions to these integrations are shown in Figure 1.  In that
figure, black lines represent the evolution of $a(t)/a_o$, red lines
represent the evolution of $\Theta$ and the blue lines represent the
evolution of the equilibrium timescale (which varies as $a^{-13/2}$)
all as a function of time, in units of the (initial) equilibrium timescale.  

Concentrating first on the slow rotators (S/L $<$1) we see two important
features.  First, in order to have significant obliquity damping the
semi-major axis is reduced substantially.  Given that we set our
initial ratio ($R_{\ast}/a_{o}$) to 0.1 any reduction of $a$ below this
value implies the planet has fallen into the star.  We see that this
occurs at approximately 2.5 T$_{eq0}$, a time when there is still
non-negligible obliquity.  Another way to look at this is to 
consider the time when
the obliquity is damped to half its original value of $\pi$/4, at this
time the equilibrium timescale has been reduced by nearly five
orders of magnitude.  One might hypothesize that we have caught the
aligned hot Jupiters in their last gasp on their death march to infall in which
their obliquity has been reduced to zero but their orbits have not
been completely exhausted.  However, the rapid decay of the
equilibrium timescale coincident with this evolution is not consistent
with the large population of aligned, yet surviving, systems that have been
observed.  In summary, even modest amounts of
obliquity damping are concurrent with significant orbital decay and a
rapidly decreasing timescale over which that orbital decay will occur.  
Similarly, for fast rotators modest obliquity damping is accompanied
by significant orbital expansion.  Therefore, in order to accomodate any substantial
obliquity evolution the planet would have to have started its orbit virtually
inside the star, an impossible scenario.  

Therefore, for both slow and
fast rotators it is impossible to get any substantial obliquity
evolution without either 1) the planet falling into the star (slow
rotators) or 2) the planet starting its orbital evolution inside the
star (fast rotators).  This long recognized problem leads to the conclusion that significant obliquity damping
can only occur if the dynamical tide is considered and the
timescale for obliquity alignment, $\tau_{10}$
is substantially less than that for orbital decay/expansion,
$\tau_{eq}$.   

\subsection{Population synthesis of obliquity evolution due to dynamical tides}
The model of \cite{alb12} assumes that hot jupiters arrived at their
current, close positions with a random distribution of obliquities as
a result of one of the dynamical processes listed in \S1.  
To mimic this scenario we start from a random distribution of obliquities and
integrate Equation (7), assuming a constant
semi-major axis, $a$ (this assumption is justified in the limit of $\tau_{10}<<\tau_{e}$).
Since the physics of tidal
dissipation is highly uncertain we run a host of models varying three
free parameters: S/L, $\tau_{10}/\tau_{eq}$ and
$\Omega_{\ast}/\Omega_{p}$.  As stated above, in order for
obliquity evolution to occur in the absence of orbital decay
$\tau_{10} < < \tau_e$, so we consider only models for which
$\tau_{10}/\tau_e < 1$ (although see comment below).  We consider
$\Omega_{ast}/\Omega_{p}$ between 0.1 and 10.  Finally, we consider values
for S/L which vary between 0.1 and 2.  Note that these are all {\it
  initial} values. %(Note that for low mass 
%planets, it is possible for $S/L >1$ and $\Omega_\ast < \Omega_p/2$).

Figure 2 shows the results of one set of our integrations (see Figure
caption for more details).  The tidal potential generated by the planet on the star results in a 
torque which depends explicitly on the angle of obliquity and goes to
zero if the obliquity is $\pi,\pi/2$ or 0.  There are
two components of this torque, one in the
direction of the spin axis of the star, the other along the orbital
axis of the planet.  Which of these components dominates the evolution
determines which eventual state of $\pi,\pi/2$ or 0 the 
planet tends to.  Hot Jupiters with $\Theta_{o}< \pi/2$ evolve toward alignment regardless 
the value of $S/L$ after $t \sim 2-10 \tau_{10}$. On the other hand, most 
hot Jupiters with $\Theta_{o} > \pi/2$ either evolve towards 
$\pi$ (in the limit of small $S/L$) or $\pi/2$ (if 
$S > L$) after $t \sim 2-10 \tau_{10}$.  Therefore, if all hot
Jupiters started with a random obliquity angle, tidal
evolution would lead to a nearly equal division between those
with prograde ($\Theta < \pi/2$) and retrograde ($\Theta
> \pi/2$) or 90$^{o}$ orbits.   The only circumstances under which all
hot Jupiters evolve to aligned systems are when $\tau_{10} \geq 0.5
\tau_{e}$.  However, as discussed in Section 3.1 such timescales would
also lead to significant orbital decay and loss of planets to their
host stars, which is inconsistent with the observations.

\subsection{Comparison with observation}

We obtain information from the website exoplanet.org and plot
the distribution of obliquity as a function of $S/L$, $\Omega_{\ast}/\Omega_{p}$ and stellar
temperature in Figure 3.  Planets around
solar type stars are represented by colored circles, while black circles 
represent planets around hot stars.  A review of these orbital properties indicates
  that very few cool stars have $S/L$ larger than 1.
  For these parameters the most likely outcome of tidal dissipation of obliquity, is that initially random distributions of obliquity evolve
  to aligned, anti-aligned or 90$^{o}$ orbits.  For $S/L \sim 0.1 -0.5$ the percentage of mis-aligned
  systems is $\sim 25-50\%$.  The observations, on the other hand,
  show an overwhelming majority of
  hot Jupiters around cool stars have aligned orbits.  The
  obliquities plotted in Figure 3 are of the projected obliquity and
  the error bars denote the likely range of true obliquity.  Despite
  projection effects, the observations are
  still inconsistent with the theoretical prediction.
  
The two limits considered, including the equilibrium tide or only
  the (1,0) dynamical
tidal, bracket the solutions.  We have avoided considering
both components simultaneously as that would require assumptions about
the efficiency of each mechanism.  However, considering the results of
3.2, that end states of obliquity evolution include a signficant
fraction of retrograde orbits, and Equation (2), that indicates
retrograde and prograde orbits may evolve on different timescales, one might argue that retrograde
systems might be preferrentially lost.  By inspection one can see that
this depends on the ratio of $\Omega_{\ast}/\Omega_{p}$.  We integrated Equations
(2)-(4) varying the initial value of $\Omega_{\ast}/\Omega_{p}$, the results of which are shown in Figure 4.  For
$\Omega_{\ast}/\Omega_{p} < 1$, where the majority of systems live,
prograde and retrograde orbits evolve on similar, or at least
indiscernible, timescales.  For $\Omega_{\ast}/\Omega_{p} \sim 1-2$,
retrograde systems could be lost, while prograde systems remain (with
high obliquity),
however this represents a small fraction of the observed systems.
Finally, for
$\Omega_{\ast}/\Omega_{p} >> 1$ retrograde orbits are
lost while prograde orbits migrate outward rapidly, inconsistent with the observed
population of close-in, aligned systems. 
Therefore, we conclude that obliquity alignment due to tidal
dissipation is unable to explain current observations.  
  
\section{Discussion}
The purpose of this paper is not to rule out the formation of 
some hot Jupiter's through dynamical processes.  The broad 
distribution of planetary eccentricity may be due to 
dynamical relaxation processes which can, in principle, lead to the
scattering of a few gas giant planets to the proximity of their host
stars.  Kozai effect may also deliver gas giants, at least in some
well known systems such as HD80606. Some of these encounters
may be sufficiently close to induce strong tidal interaction between
the planets and their host stars.  It is entirely possible that dynamical processes may have led to the 
formation of some hot Jupiter with diverse obliquities around both 
solar type and hotter main sequence stars.  

%For example, significant envelope 
%losses during close stellar passage may lead to the ejection of 
%coreless gas giants (Guillochon et al 2012) or the transformation 
%from gas giants with cores to downsized super Earths (Liu et al 2013).  

However, the dynamical model alone cannot account for the dichotomy between 
the obliquity distribution of hot Jupiters around solar type 
and intermediate-mass stars because this mechanism does not 
depend on the mass of the host stars.  \cite{alb12} attribute
this difference to the efficient dissipation of hot Jupiters' tidal 
perturbation on their solar type host stars whereas that process
is inefficient in intermediate mass stars.  The results 
presented here pose a challenge to this tidal alignment model.

We have shown that if one considers only the equilibrium tide dissipation leads to
such severe orbital decay that any systems which have their obliquity
aligned fall into their host star, or, for fast rotators, will have their orbits expand so
drastically that they would have to have started their orbit inside
the star.  At the other extreme, if we instead consider the dynamical tide proposed by
\cite{lai12} tidal dissipation of obliquity produces both prograde
and retrograde systems for a population of hot Jupiters with random initial
$\Theta$.  We confirmed that although hot
Jupiters with initially prograde obliquities (with $\Theta 
< \pi/2$) would become aligned, those with initially 
retrograde obliquities (with $\Theta > \pi/2$) would attain
either anti-aligned or orthogonal obliquity.   Such an 
obliquity distribution is inconsistent with that observed for hot
Jupiters around solar type stars.  We found that the possibility that
these systems evolve by the dynamical tide but that retrograde orbits
are preferentially damped can only explain systems in which
$\Omega_{\ast}/\Omega_{p} \sim 1$ and there are few of these
systems.  Therefore, this also, can not explain the current
observations. 

In a series of papers (Rogers et al 2012, 2013), we proposed
an alternative model for the observed dichotomy between the
spin-orbit alignment for hot Jupiters around solar-type and
intermediate mass stars.  We suggest that most hot Jupiters
migrated to the proximity of their host stars through type II
migration (Lin ei al. 1996).  These planets retained the 
angular momentum vector associated with the disk.  But the spins of
their hot host stars may be modulated due to the excitation, propagation, and 
dissipation of internal gravity waves, a process which is only
efficient in hot stars.  To date our gravity wave model is the only scenario which can provide a natural
explanation of the observed difference between the $\Theta$
distribution of hot Jupiters around solar type stars and that
around intermediate-mass stars.  Furthermore, with this theory, there
is no need to introduce multiple scenarios to account for the migration 
of hot Jupiters versus that of multiple-planet systems as 
suggested by \cite{alb13}.

\acknowledgments
We thank Dong Lai, Yanqin Wu, Peter Goldreich and an anonymous referee
for useful
conversations.  Support for this work was provided by NASA grant
NNG06DGD44G. T.M. Rogers is supported by NSF ATM Faculty Position
in Solar Physics under award number 0457631.  D.N.C. Lin was supported
by NASA (NNX08AM84G), NSF (AST-0908807) and UC Lab Fee. 
 
%\bibliographystyle{apj}

%\bibliographystyle{apj}
%\input{hotjupob.bbl}

%\bibliography{hotjupob}
%
\begin{figure}
%\epsscale{1.00}
%\plotone{sma-obliqevol-eqt}
%\plotone{fig1.eps}
\includegraphics[width=4in]{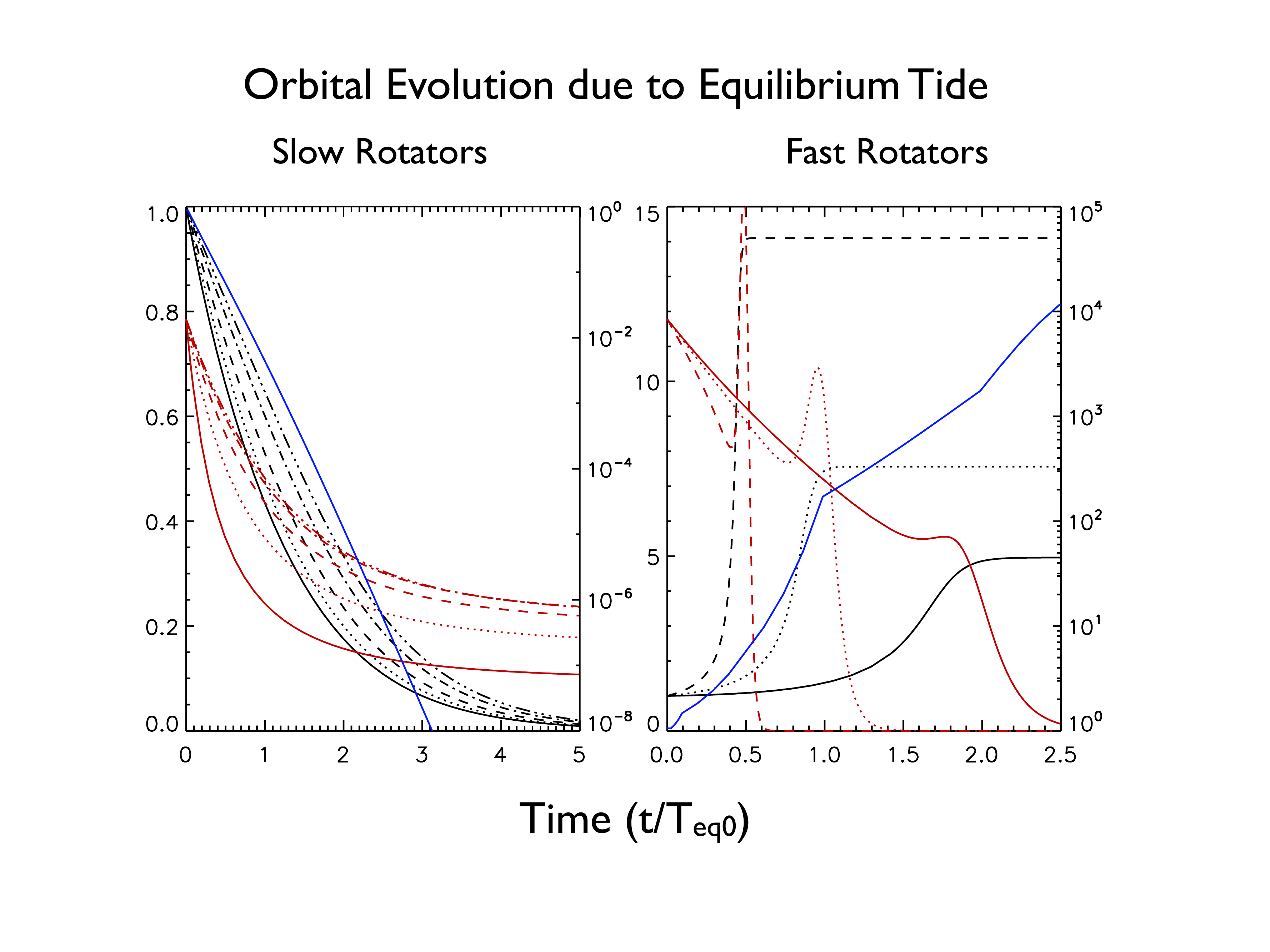}
\caption{Evolution of semi-major axis, a (black lines), and obliquity
  $\Theta$ (red lines) as a function of time for the equilibrium
  tide.  Left axis represents semi-major axis and obliquity, right
  axis represents equilibrium timescale.  Left hand panel show slow rotators (S/L$<$1), while the right
  hand panel shows fast rotators (S/L$>$1).  $\Theta$ was initially set to
45$^{o}$ in these integrations.  Different linetypes represent
different values of S/L.  }
\end{figure}

\begin{figure}
%\epsscale{1.00}
%\plotone{obliqevol-f-hdt}
%\plotone{fig2.eps}
\includegraphics[width=4in]{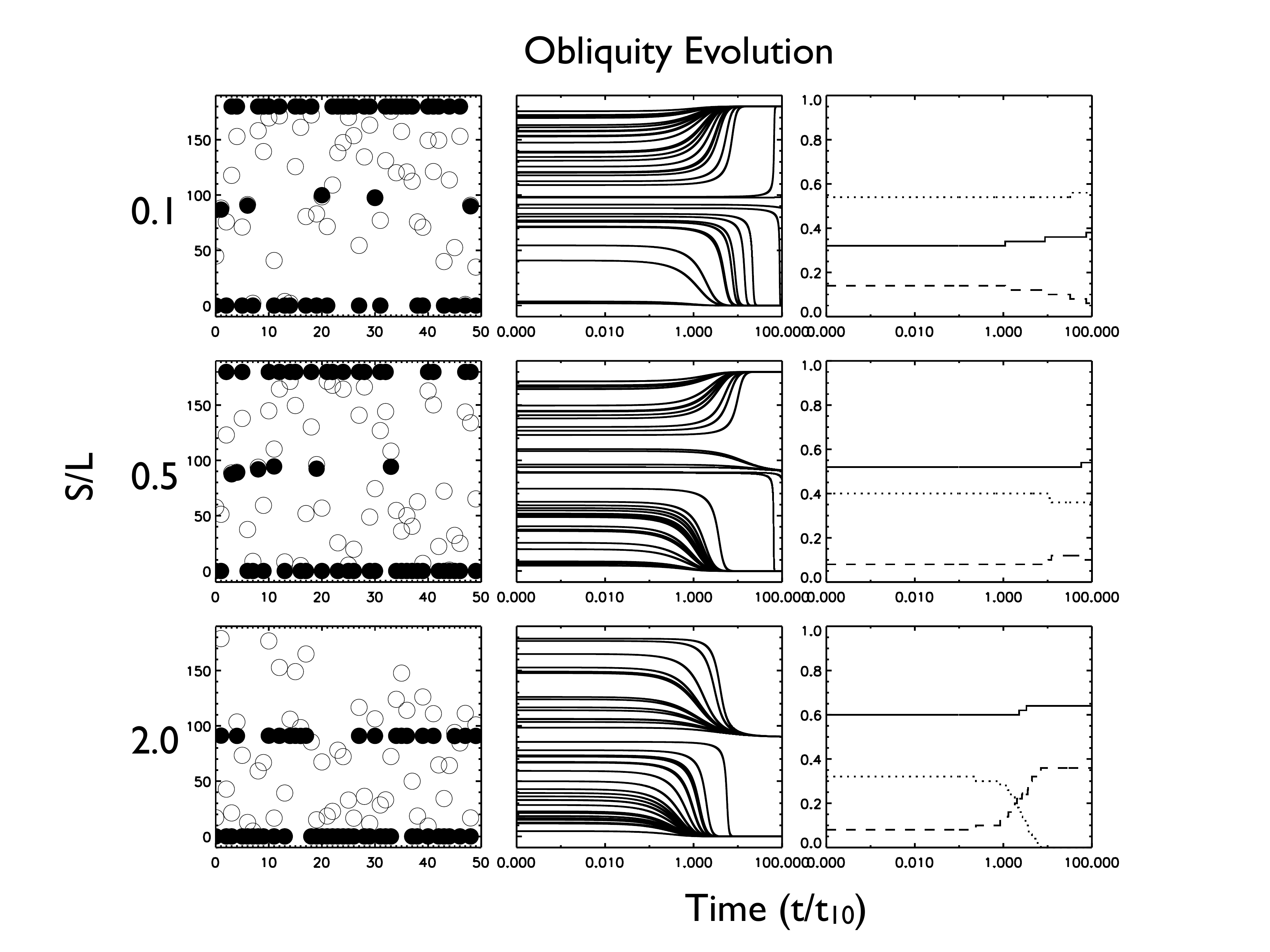}
\caption{Obliquity evolution of a population of objects with initially
  random obliquity and with $\tau_{10}/\tau_{eq} =0.001$.  Left panels show the original random distribution
  of obliquities for the 50 objects (open circles), while filled
  circles show the distribution
  of obliquity after 30$\tau_{eq}$.  Middle
  column shows the time evolution of the obliquity, while the right
  panel shows the fraction of objects as a function of time with
  prograde orbits (solid line), retrograde orbits (dotted line) and
  90$^{o}$ orbits (dashed line).  In all scenarios the objects evolve
  to prograde, retrograde or 90$^{o}$ orbits.  Note that since we are
  keeping a fixed, S/L does not evolve.}
\end{figure}

\begin{figure}
%\epsscale{1.00}
%\plotone{fig3n}
%\plotone{fig3.eps}
\includegraphics[width=4in]{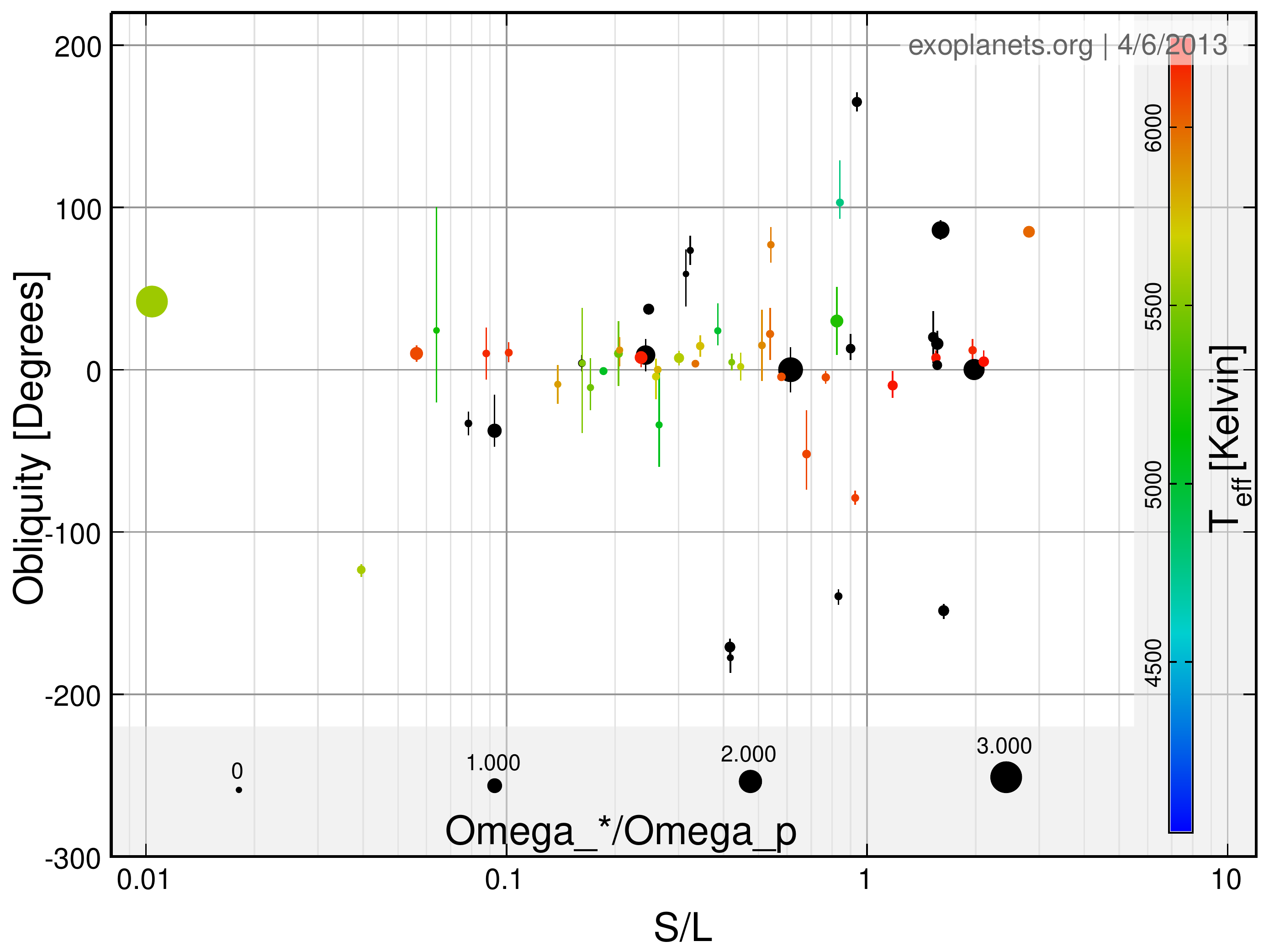}
\caption{Distribution of projected obliquity as a function of temperature, S/L
  and $\Omega_{\ast}/\Omega_{p}$.  Cool stars
  are represented in color, while hot stars are represented in black.}
\end{figure}

\begin{figure}
%\epsscale{1.00}
%\plotone{fig3n}
%\plotone{fig3.eps}
\includegraphics[width=4in]{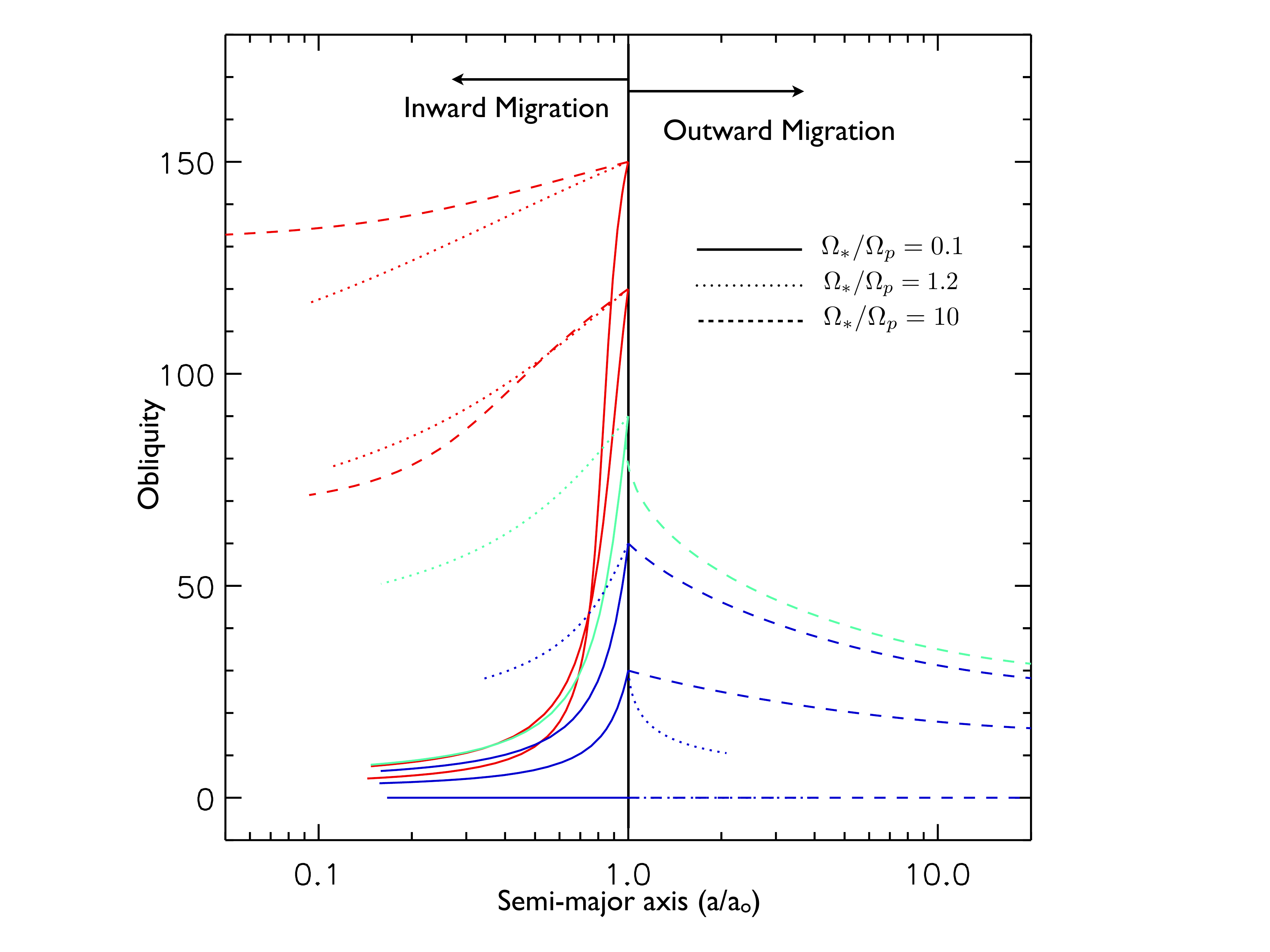}
\caption{Evolution of obliquity and semi-major axis for various ${\it
    initial}$ values
of $\Omega_{\ast}/\Omega_{p}$, after 2$T_{eq0}$.  Red, blue and green lines represent initially
retrograde, prograde and 90$^{o}$ orbits, respectively}
\end{figure}

\end{document}